\documentclass[review]{elsarticle}

\usepackage{lineno,hyperref,epstopdf}
\usepackage{booktabs,amssymb}
\usepackage{graphicx,amsmath,float}
\usepackage{multirow,diagbox,makecell,rotating}
\modulolinenumbers[5]

\date{}
\begin{document}

\begin{frontmatter}
\title{Information Diffusion Prediction with Latent Factor Disentanglement}



\author[mymainaddress]{Haoran Wang}
\ead{wanghaoran@bupt.edu.cn}

\author[mymainaddress]{Cheng Yang\corref{mycorrespondingauthor}}
\cortext[mycorrespondingauthor]{Corresponding author}
\ead{yangcheng@bupt.edu.cn}


\address[mymainaddress]{School of Computer Science, Beijing University of Posts and Telecommunications, China}

\begin{abstract}
Information diffusion prediction is a fundamental task which forecasts how an information item will spread among users. In recent years, deep learning based methods, especially those based on recurrent neural networks (RNNs), have achieved promising results on this task by treating infected users as sequential data. However, existing methods represent all previously infected users by a single vector and could fail to encode all necessary information for future predictions due to the mode collapse problem. To address this problem, we propose to employ the idea of disentangled representation learning, which aims to extract multiple latent factors representing different aspects of the data, for modeling the information diffusion process. Specifically, we employ a sequential attention module and a disentangled attention module to better aggregate the history information and disentangle the latent factors. Experimental results on three real-world datasets show that the proposed model SIDDA significantly outperforms state-of-the-art baseline methods by up to $14\%$ in terms of $hits@N$ metric, which demonstrates the effectiveness of our method.
\end{abstract}


\end{frontmatter}


\section{Introduction}
The phenomenon of information diffusion, dubbed as information cascade, is ubiquitous in our daily lives, \textit{e.g.}, the spread of breaking news or virus. Information diffusion prediction is an important and challenging task, which aims at forecasting the future properties or behaviors of an information cascade, such as the eventual size~\cite{deepcas-macro,cascn-macro,macro-coupledGnn} or the next infected user~\cite{NDM,DCE-micro,infVAE-micro}. Current studies of information diffusion prediction have been successfully adopted for many real-world scenarios including epidemiology~\cite{wallinga2004different}, viral marketing~\cite{leskovec2007dynamics}, media advertising~\cite{li2013popularity} and the spread of news and memes~\cite{leskovec2009meme,vosoughi2018spread}.

\begin{figure}[htb]
\centering
\includegraphics[scale=0.4]{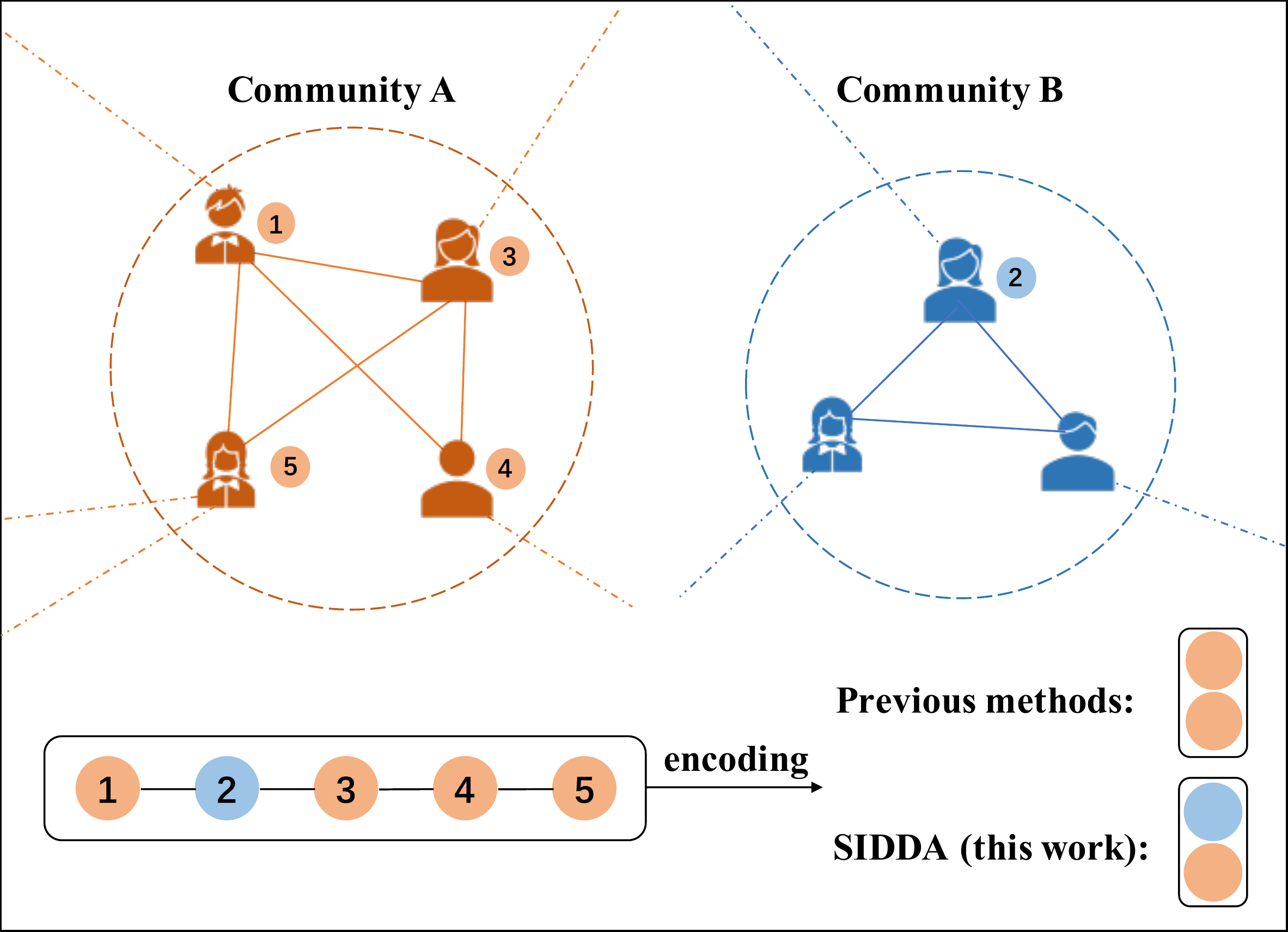}
\caption{An illustration of mode collapse problem in information cascade modeling. }
\label{fig_intro}
\end{figure}

During the last decade, deep learning techniques have shown their effectiveness in computer vision~\cite{krizhevsky2017imagenet} and natural language processing areas~\cite{devlin2018bert}. Recently, methods~\cite{deepcas-macro,CYAN-RNN,TopoLSTM,SNIDSA} based on recurrent neural networks (RNNs) also achieved promising results in information cascade modeling by treating influenced users as sequential data ranked by their infection timestamps. In RNNs, the entire information diffusion history is encoded into a real-valued vector as the hidden state. However, representing all previously infected users by a single vector could fail to encode all necessary information for future predictions due to the mode collapse~\cite{che2016mode} problem. As illustrated in Fig.\ref{fig_intro}, an information item is spreading through two communities and the cascade sequence has infected $5$ users ($4$ from community A and $1$ from community B). An information cascade model need to encode these users. Learned representations of conventional methods can encode the most important factor (community A) for future predictions, but fail to remember others (community B).

To address this problem, we propose to employ the idea of disentangled representation learning, which aims to extract multiple latent factors representing different aspects of the data, for modeling the diffusion history of an information cascade. Though disentangled representation learning was originally proposed in controllable image generation~\cite{chen2016infogan}, it has been successfully adopted for other areas such as text generation~\cite{stylicSPG-disentangled} as well. To the best of our knowledge, we are the first to use disentangled representation learning for information diffusion prediction. 

To be more specific, we propose a novel Sequential Information Diffusion model with Disentangled Attention (SIDDA) by applying a sequential attention module and a disentangled attention module on RNNs. The former module can help better aggregate the history information and the latter one will disentangle the hidden state vector into multiple representations. We further employ Gumbel-Softmax~\cite{gumbel} technique to encourage bigger differences between the disentangled representations. Consequently, we are able to learn multiple hidden state vectors characterizing different latent factors for future predictions at each step of cascade modeling. Different hidden state vectors can complement each other and provide more comprehensive information to predict the next infected users.

We conduct experiments on three public real-world datasets and employ the task of next infected user prediction for evaluation. Experimental results show that the proposed model SIDDA significantly outperforms state-of-the-art baseline methods by up to $14\%$ in terms of $hits@50$ metric.

Our contributions are as follows:
\begin{itemize}
    \item To the best of our knowledge, we are the first work to adopt the idea of disentangled representation learning for information cascade modeling.
    \item We propose SIDDA, a novel method which can learn disentangled representations encoding different factors for information diffusion prediction.  
    \item Experiments on three real-world datasets demonstrate that our proposed method outperforms state-of-the-art baselines significantly.
\end{itemize}

\section{Related Work}
\subsection{Information Diffusion Prediction}
There are two kinds of tasks associated with information diffusion prediction: (1) Studying the growth of cascades and (2) Predicting the next activated node in the cascade. In some recent works~\cite{cascn-macro,NDM,FOREST,zhou2020survey}, these tasks are classified as macroscopic and microscopic predictions. 

\subsubsection{Macroscopic Information Diffusion Prediction}
Macroscopic information diffusion prediction, also known as popularity prediction, aims at predicting cascade sizes in the future~\cite{zhao2015seismic}. Previously, related works are based on feature-based approaches~\cite{cancascadebepredicted} and generative approaches~\cite{generative-hawkes}. In recent years, with the success of deep learning, methods based on Recurrent Neural Networks (RNNs) are proposed, \textit{e.g.}, DeepCas~\cite{deepcas-macro} and Deephawkes~\cite{deephawkes-macro}. Some researchers also introduce Graph Neural Networks (GNNs) to model the underlying social graph and diffusion paths, \textit{e.g.}, CoupledGNN~\cite{macro-coupledGnn} and HDGNN~\cite{heterogeneous4-HDGNN-macro}. 

\subsubsection{Microscopic Information Diffusion Prediction}
Our work focuses on microscopic information diffusion prediction, which aims to forecast the next activated node given previously infected users in an information cascade. 


With the development of deep learning, various kinds of neural networks have been adopted for microscopic information diffusion prediction. Generally, RNN, GNN and attention mechanism are widely used, and give promising results. Topo-LSTM~\cite{TopoLSTM} proposes a novel LSTM to model tree-structured cascades. CYAN-RNN~\cite{CYAN-RNN} and Deep-Diffuse~\cite{deepdiffuse} take timestamps into consideration with the help of temporal point processes. DyHGCN~\cite{heterogeneous2-DyHGCN} proposes a heterogeneous graph convolutional network to model the social graph and dynamic diffusion graph jointly. HDD~\cite{heterogeneous1} exploits meta-path in the diffusion graph and learns heterogeneous network representations using GNN. There are also some models fully based on attention mechanisms, \textit{e.g.}, DAN~\cite{DAN}, Hi-DAN~\cite{Hi-DAN} and NDM~\cite{NDM}.

Besides, some works target on both macroscopic and microscopic information diffusion predictions. FOREST~\cite{FOREST} makes use of reinforcement learning to incorporate the ability of macroscopic prediction. DMT-LIC~\cite{dmt-lic-related} is a multi-task learning framework with a shared-representation layer and two different task layers for predictions.

However, existing methods represent previously infected users by a single vector and could encounter the mode collapse problem, which would fail to encode all necessary information for future predictions.
\subsection{Disentangled Representation Learning}
Disentangled representation learning which aims to learn representations of different latent factors hidden in the observed data~\cite{representationlearning,chen2016infogan}, has been successfully used in various fields. In the field of recommendation systems, a few works are proposed to learn disentangled item representations~\cite{ma2019learningcosinedot}. \cite{ma2020disentangled} performs self-supervision in the latent space, getting disentangled intentions from item sequences for product recommendations. In the field of natural language processing,
SPG~\cite{stylicSPG-disentangled} proposes an unsupervised way to generate stylistic poetry by maximizing the mutual information between representations and generated text. GNUD~\cite{GNUD-disentangled} explores user interest disentanglement in news recommendation by making use of DisenGCN~\cite{madisentangledgnn-DisenGCN}. Besides, disentangled representation learning is also successfully used in the modeling of graph-structured data~\cite{multi-aspect,madisentangledgnn-DisenGCN, disentangledGNN-IPGDN}.

As far as we know, we are the first to adopt the idea of disentangled representation learning for information diffusion prediction.

\section{Method}
In this section, we will formalize the problem of information diffusion prediction and introduce the notations. Then we will propose an RNN-based information diffusion prediction model and a novel attention mechanism to learn disentangled representations. Finally, we will predict the next infected node using the disentangled representations.
\subsection{Problem Definition}
Given a node (or user) set $V$ and a cascade set $C$, we have $V=\{v_1,v_2,...,v_N\}$ and $C=\{c_1,c_2,...,c_M\}$. Here $N$ is the size of the node set, and $M$ is the number of cascades. Each cascade $c_i \in C$ spreading among users is a sequence of nodes $\{v^i_1,v^i_2,...,v^i_{\left | c_i \right |}\}$, ordered by their activation timestamps. Following the same setting in previous works~\cite{NDM, TopoLSTM, SNIDSA, FOREST}, we only keep the order of nodes and ignore the exact timestamps. A detailed modeling of timestamp information will be left for future work.

Information diffusion prediction can be formulated as: given the cascade sequence of previously activated nodes $\{v^i_1,v^i_2,...,v^i_t\}$ in cascade $c_i$ for $t=1,2,...,\left | c_i \right |-1$, predicting the next activated node $v^i_{t+1}$. In other words, our goal is to build a model that is able to learn the conditional probability function $p(v^i_{t+1}|\{v^i_1,v^i_2,...,v^i_t\})$ for each cascade $c_i$ where $1\leq i \leq M$. We will focus on the modeling of a single cascade and omit the superscript for simplification in the rest of the paper.

\subsection{Model Architecture}
\begin{figure}[H]
\centering
\includegraphics[scale=0.4]{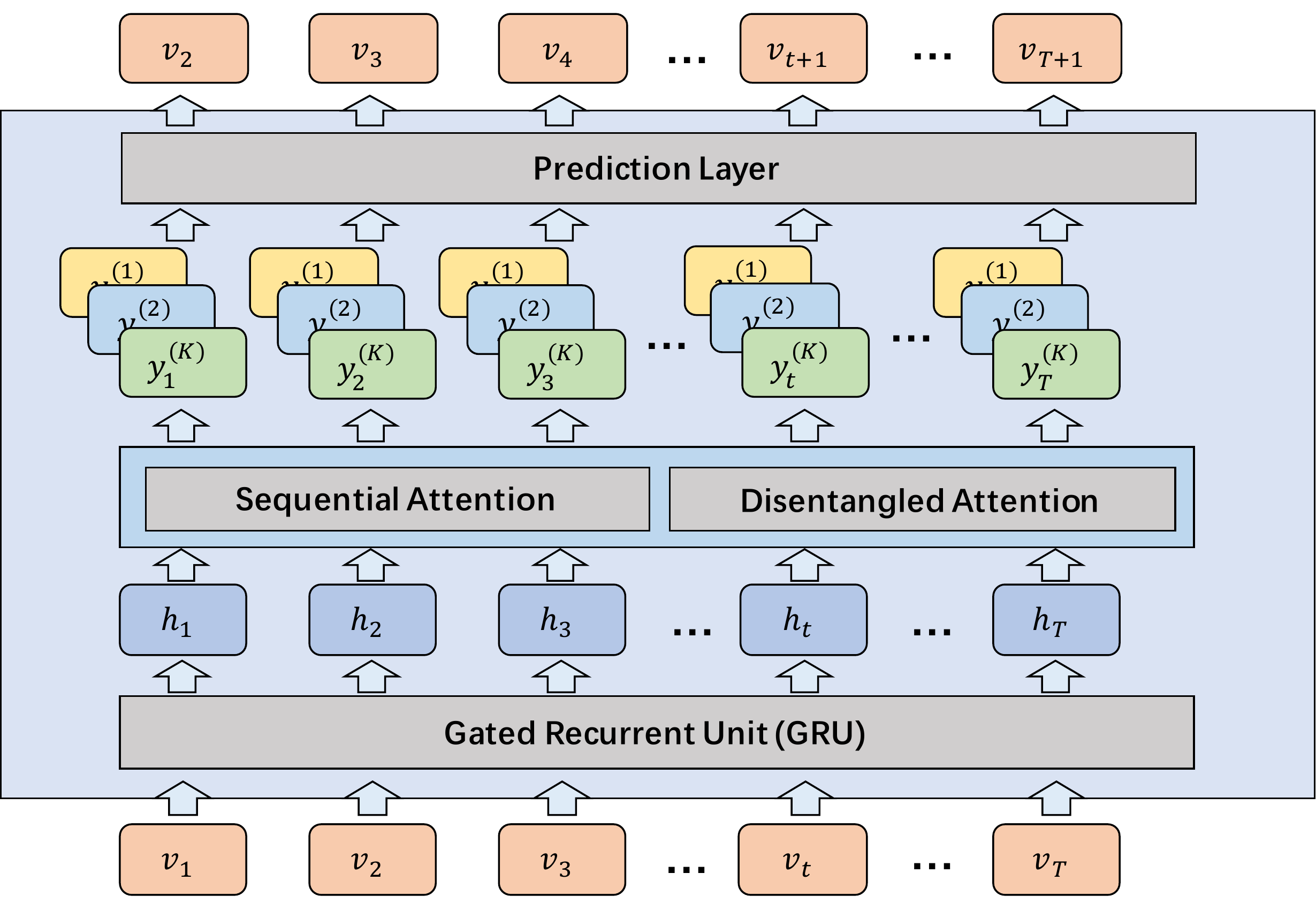}
\caption{An illustration of our proposed model SIDDA. The model will derive $K$ disentangled representations $\mathbf{y_t^{(k)}}(1 \leq k \leq K)$ at time $t$, with the help of a disentangled attention module.}

\label{model_overall}
\end{figure}

The framework of our method is shown in Fig.\ref{model_overall}. Firstly, we encode every node into node embedding by an embedding-lookup layer. Then we employ a Gated Recurrent Unit (GRU) as the basis to model the node sequence. Given the infected node sequence, the GRU model can output a hidden state representing the sequential history at each time step. Then a sequential attention module and a disentangled attention module are applied to better aggregate the history information and disentangle the hidden state vector into multiple representations. Finally, disentangled representations will be used for predicting the next infected node.

\subsubsection{Gated Recurrent Unit}
Recurrent neural networks(RNNs) have shown great potentials in modeling sequence data. Previous works~\cite{TopoLSTM,SNIDSA,FOREST,CYAN-RNN,deepdiffuse,lstm-cnn-hin-kbs} used RNNs as their basis to model information diffusion cascades. Firstly, we represent each node as a vector $\mathbf{x}\in \mathbb{R}^D$. By feeding the activated node embedding $\mathbf{x_t}$ into an RNN at every timestamp, we can have a hidden state $\mathbf{h_t}$ capturing the history information of all previously activated nodes as the output.

Gated Recurrent Unit (GRU) can alleviate the vanishing gradient problem compared with a standard RNN. It can be considered as a variation of the LSTM but is computationally cheaper. Therefore, we select GRU as our model basis. 

\begin{equation}
\mathbf{h_t} = GRU(\mathbf{h_{t-1}}, \mathbf{x_t})
\label{gru}
\end{equation}

\subsubsection{Sequential Attention Module}
Note that every hidden state $\mathbf{h_t}$ in GRU contains information about the input node sequence with a focus around position $t$. To aggregate the sequential history information of all positions automatically, we propose to employ attention mechanism, which was originally proposed in neural machine translation~\cite{attentionpaper}, and learn to assign different attention weights to previous hidden states $\mathbf{h_i}$ $(i\leq t)$. Formally, the sequential attention weight $\alpha_{it}$ of the $i$-th hidden state is computed by:
\begin{gather}
\alpha_{it} = \frac{exp(\frac{1}{\sqrt{D}} e_{it} )}{\sum_{j=1}^{t}exp(\frac{1}{\sqrt{D}}e_{jt})} \\
e_{it} = d(\mathbf{h_i},\mathbf{h_t})\notag
\end{gather}
where $\frac{1}{\sqrt{D}}$ is a scaling coefficient and $h_i\in \mathbb{R}^D$ is the $i$-th hidden state. $d(\cdot,\cdot)$ is a distance function, and we use dot product in this work.

\subsubsection{Disentangled Attention Module}
Inspired by the recent advances in disentangled representation learning~\cite{madisentangledgnn-DisenGCN,ma2020disentangled}, we propose our disentangled attention module to learn multiple representations characterizing different latent factors for future predictions at each step of cascade modeling. 

We assume that there are $K$ main latent factors that affect the diffusion of an information item. Take Twitter as an example, an information item may spread through $K$ different communities or users are infected by tweets from $K$ topics. We want to disentangle these diverse latent factors from the hidden state $\mathbf{h_t}$ to get the disentangled representations.


\begin{figure}[H]
\centering
\includegraphics[scale=0.4]{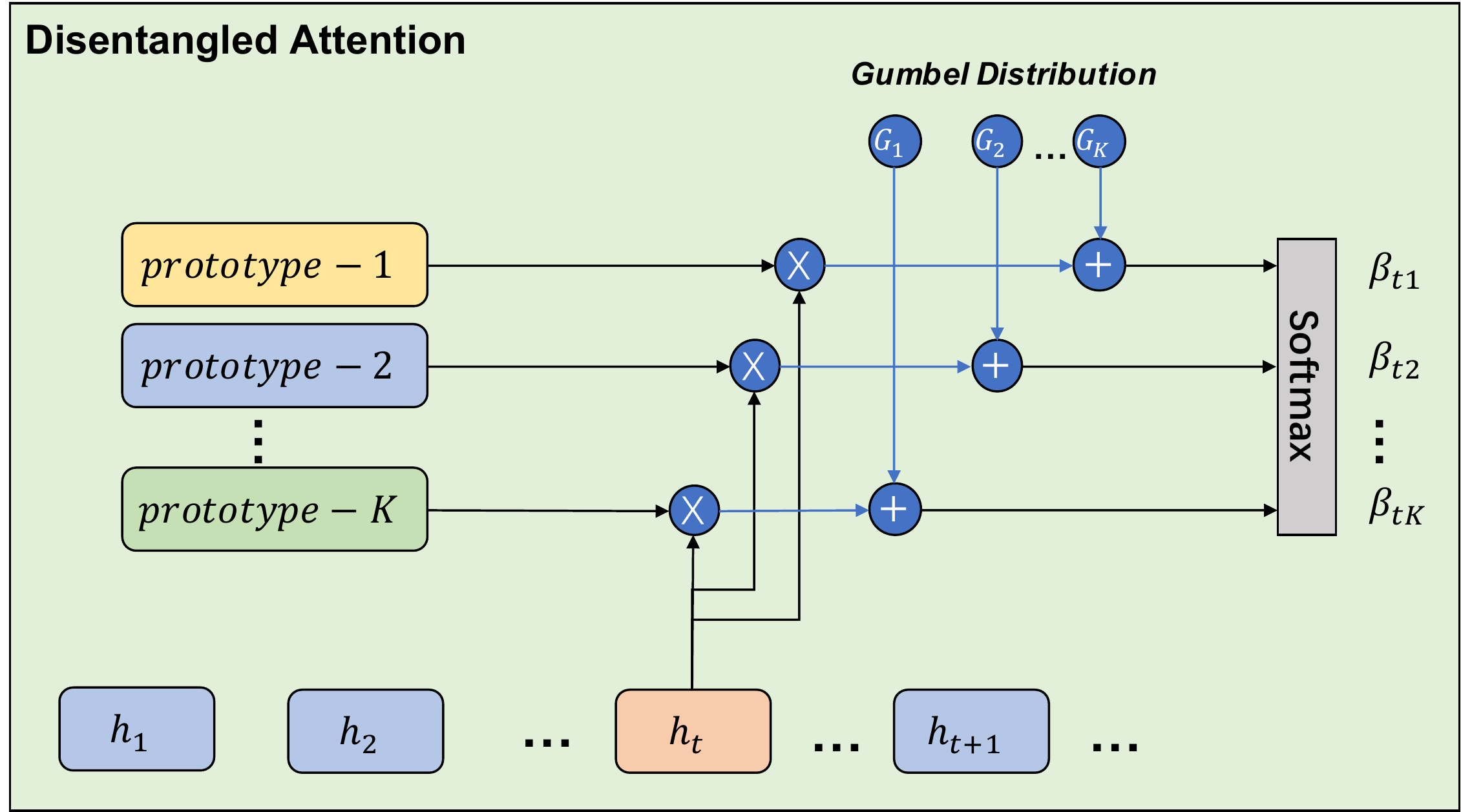}
\caption{An illustration of the disentangled attention module.}
\label{model_pki}
\end{figure}


To be more specific, we set $K$ prototype embeddings $\mathbf{p_k}$ $(1\leq k\leq K)$ to represent the $K$ potential factors. We expect these prototype embeddings can capture diverse properties after training. Then for each hidden state $h_i$, we will compute its similarity with $K$ prototypes by the distance between $\mathbf{h_i}$ and $\mathbf{P}$. The equations are as follows:
\begin{equation}
\begin{gathered}
\beta_{ik} = \frac{exp(\frac{1}{\sqrt{D}}d(\mathbf{h_{i}},\mathbf{p_k}))}{\sum_{{k}'=1}^{K}exp(\frac{1}{\sqrt{D}}d(\mathbf{h_{i}},\mathbf{p_{{k}'}}))} \\
d(\mathbf{h_i},\mathbf{p_k}) = \frac{\mathbf{h_i}\cdot \mathbf{p_k}}{\|{\mathbf{h_i}}\| \|{\mathbf{p_k}}\|} 
\label{pki}
\end{gathered}
\end{equation}
where $k=1,2,...,K$ and $i=1,2,...,t$. $\{\mathbf{p_k}\in \mathbb{R}^D:1\leq k\leq K\}$ are prototype embeddings of $K$ latent factors. $d(\cdot,\cdot)$ is the distance function, and we select cosine-similarity instead of dot product here because the latter is more vulnerable when it comes to mode collapse~\cite{ma2019learningcosinedot}. 

Then we apply a Softmax function to compute the attention score $\beta_{ik}$ for further usage. It is worth noting that by using Softmax, all of the $\beta_{ik}$ $(1\leq k\leq K)$ tend to have a similar result. The hidden state $h_i$ will be disentangled to $K$ latent factors equally, which would harm the performance of disentanglement. We further employ Gumbel-Softmax to alleviate this drawback and enhance the disentanglement effect. 


Finally, we can combine the sequence attention and disentanglement attention together, using $\alpha_{ij}$ and $\beta_{ik}$ to draw the weighted sum result $y_i^{(k)}$ to represent the hidden state under the $k$-th factor:
\begin{equation}
\mathbf{y_t^{(k)}} = LayerNorm(\sum_{i=1}^{t}\alpha_{it}\cdot \beta_{ik}\cdot \mathbf{h_i})
\label{pkipi}
\end{equation}
where $k=1,2,...,K$. When $K=1$, our method degenerates into a GRU model with an attention mechanism. We name our model as Sequential Information Diffusion model with Disentangled Attention (SIDDA). We further apply Gumbel-Softmax trick~\cite{gumbel} to encourage the disentanglement.

\subsection{Optimization Objective}
We can get $K$ disentangled representations at each timestamp $t$ using Eq. (\ref{pkipi}). Different hidden state vectors can complement each other and provide more comprehensive information to predict the next infected users. Therefore, we compute the similarity, \textit{i.e.}, dot product, between these $K$ disentangled representations $\mathbf{y_t^{(k)}}$ and node embeddings $\mathbf{X}$ to predict the next activated node. Hence, the loss function can be defined as follows:
\begin{equation}
\begin{aligned}
L(\mathbf{\theta},t) &= -\log p_{\mathbb{\theta}}(\mathbf{x_{t+1}}|\mathbf{y_t}) \\
&= -\log \frac{\max_{k\in \{1,2,...,K\}}exp(\frac{1}{\sqrt{D}} \mathbf{x_{t+1}}\cdot \mathbf{y_t^{(k)}} )}
{\sum_{{v}' \in V}\max_{k\in \{1,2,...,K\}}exp(\frac{1}{\sqrt{D}} \mathbf{x_{{v}'}}\cdot \mathbf{y_t^{(k)}})}
\label{loss}
\end{aligned}
\end{equation}

It is noteworthy that in Eq. (\ref{loss}), the Softmax function is applied to all of the $K$ disentangled factors. The model is forced to preserve different information in $\mathbf{y_t^{(k)}}(1\leq k \leq K)$. Therefore, we do not need to use regularization terms to restrict the distribution of $\mathbf{y_t^{(k)}}$ in our model.

\section{Experiments}
In this section, we will introduce the datasets used in our experiments, and the state-of-the-art baselines which will be compared with our proposed method. We further introduce the evaluation metrics used to evaluate the performance of the baselines and our method. We will design comparative experiments and ablation experiments to show the superiority of our method.
\subsection{Datasets}
\textbf{Twitter}~\cite{twitterdata} dataset collects tweets published in October 2010, containing URLs in the message body. Each URL is considered to be a unique marker of information, spreading among users.

\textbf{Douban}~\cite{doubandata} is a Chinese social networking service network where users can share content about books. In our experiment, we consider each book as an information item, and a user is activated if the user reads the book. 

\textbf{Memetracker}~\cite{memetrackerdata} contains millions of online mainstream social media activity, tracking the most frequent phrases, \textit{i.e.}, memes. Memes were information items being shared among websites. 

\begin{table}[htbp]
	\centering
	\caption{Statistics of datasets.}
	\begin{tabular}{ccccc}
		\toprule  
		Dataset&\# Cascades&\# Nodes&\# Links &Average Length \\ 
		\midrule  
		Twitter&3,442&12,627&309,631&32.60\\
		Douban&10,602&23,123&348,280&27.14\\
		Memetracker&12,661&4,709&-&16.24\\
		\bottomrule  
	\end{tabular}
	\label{dataset}
\end{table}

We follow the experiment setting in~\cite{FOREST}, randomly selecting 80\% of cascades for training, 10\% for
validation, and 10\% for test. The statistics of datasets
are listed in Table \ref{dataset}. Twitter and Douban datasets have an underlying social graph, which will be used by some of the baseline models.
\subsection{Baselines and Experimental Settings}
We compared our method with several state-of-the-art baselines.

\textbf{DeepDiffuse}~\cite{deepdiffuse} is an LSTM based model with an attention mechanism, utilizing node sequence and their corresponding activated time. We replace timestamps with integer activated steps in the cascade.

\textbf{Topo-LSTM}~\cite{TopoLSTM} is a topological recurrent network, which can model diffusion topology as dynamic directed acyclic graphs (DAGs). In our experiment, the cascade structure is extracted from the underlying social graph.

\textbf{NDM}~\cite{NDM} employs convolutional network and attention mechanism for cascade modeling. It makes relax assumptions on the datasets and doesn't need a diffusion graph.   

\textbf{SNIDSA}~\cite{SNIDSA} is a novel sequential neural network with structure attention. It utilizes both the sequential nature of a cascade and structural characteristics of the underlying social graph with the help of a gating mechanism. 

\textbf{FOREST}~\cite{FOREST} is a novel GRU-based information diffusion model. It extracts underlying social graph information and uses reinforcement learning to incorporate macroscopic prediction ability into itself. 

\subsection{Evaluation Metrics and Experiment Setting}
 As pointed out by~\cite{TopoLSTM}, there can be an arbitrary number of potential candidates, information diffusion prediction can be seen as a retrieval task. We use two popular information retrieval evaluation method $\textbf{hits}@N$ and $\textbf{map}@N$. The same evaluation metrics are also used~\cite{FOREST,CYAN-RNN,deepdiffuse, Hi-DAN}. We set $N=10,50,100$ for evaluation. Since SNIDSA and TopoLSTM need an underlying social graph in the dataset, we exclude them for Memetracker. 
 
 We implement our method in PyTorch. We use Adam optimizer for mini-batch gradient descent and use dynamic learning rate. The dropout rate is set to $1e-4$. The temperature parameter of Gumbel-Softmax $\tau=1.0$. 
 
\begin{figure}[htbp]

\centering
\begin{minipage}[t]{0.49\textwidth}
\includegraphics[width=\textwidth]{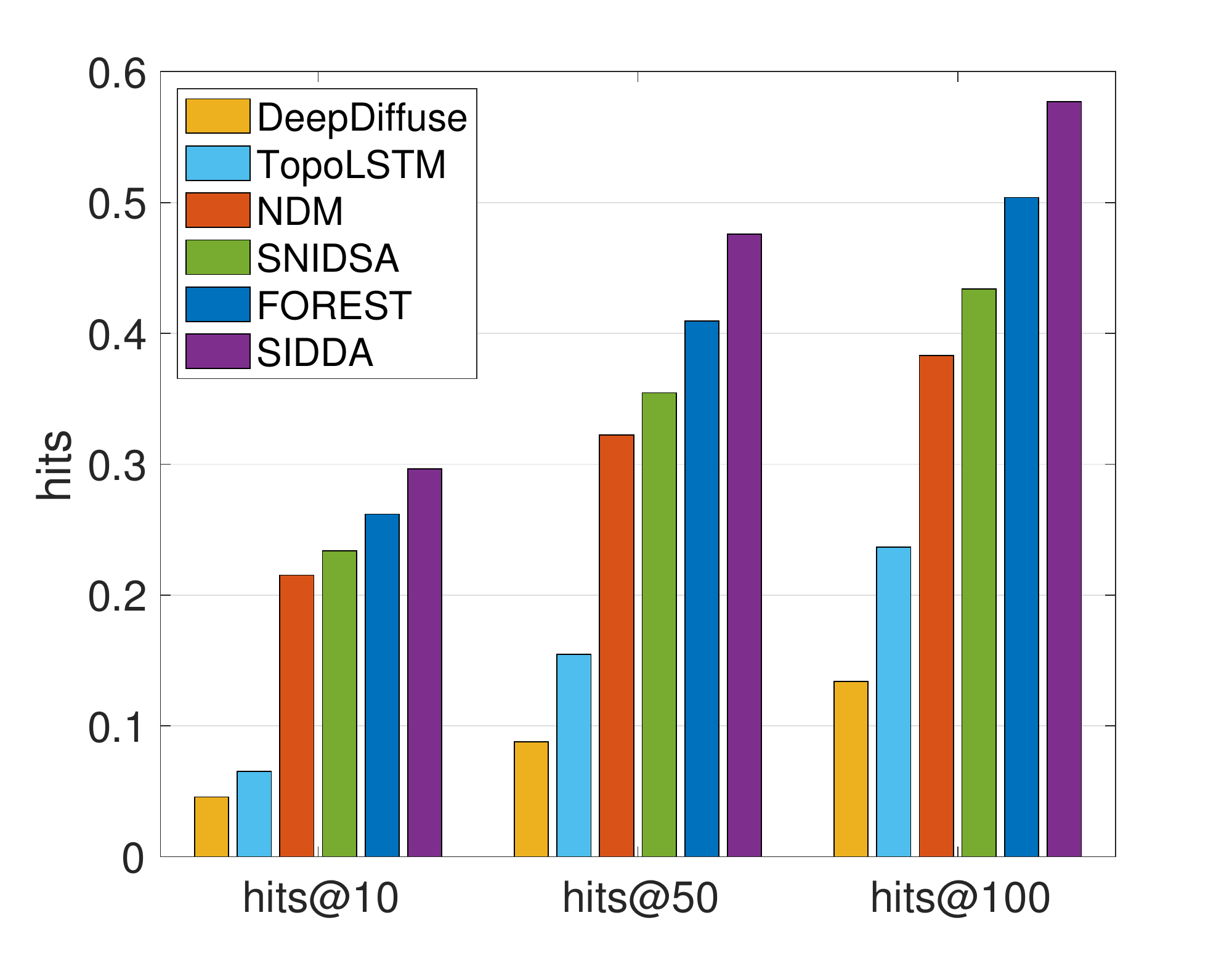}
\vspace*{-12mm}
\caption{$hits@N$ of Twitter}
\label{overall1}
\end{minipage}
\begin{minipage}[t]{0.49\textwidth}
\includegraphics[width=\textwidth]{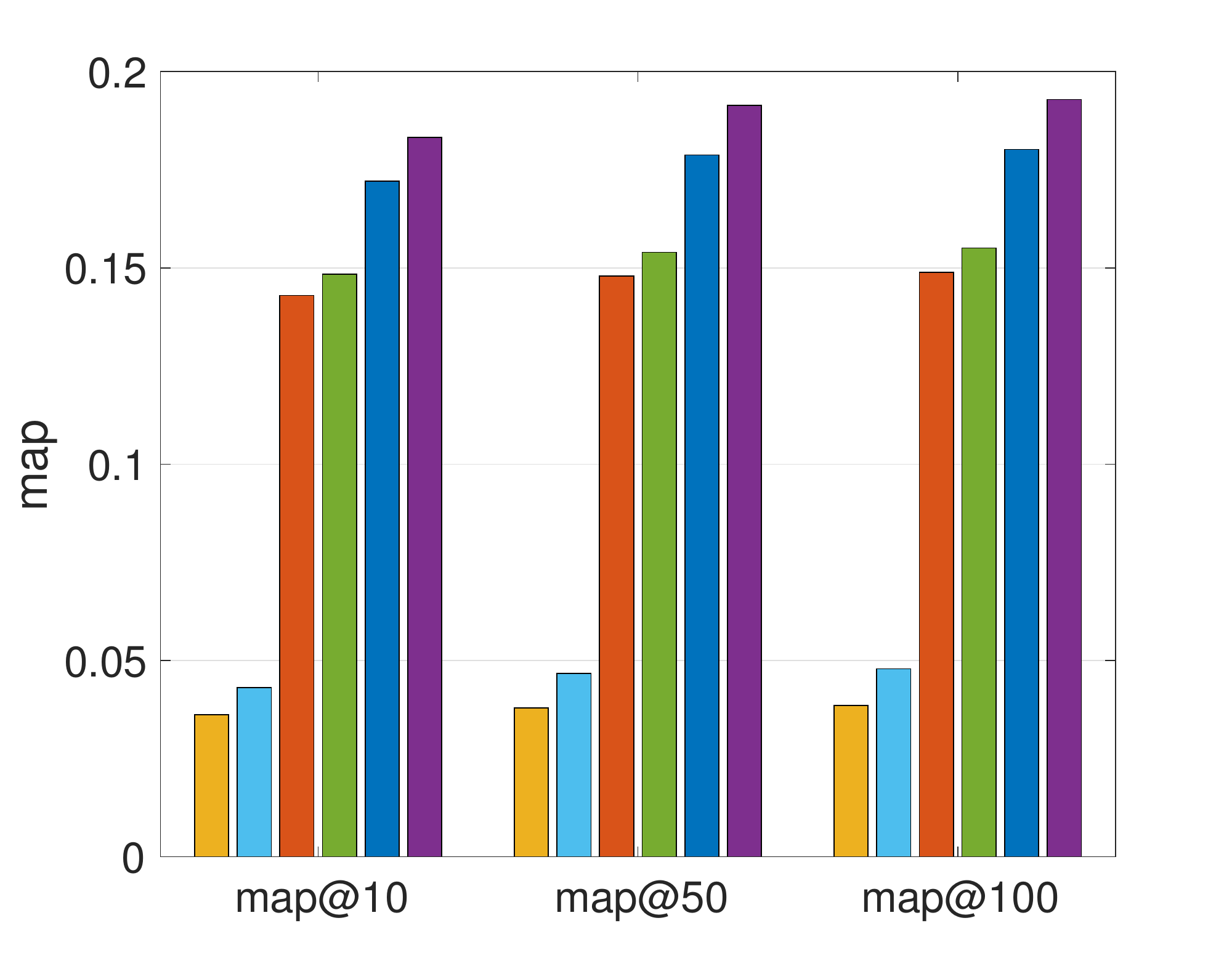}
\vspace*{-12mm}
\caption{$map@N$ of Twitter}
\label{overall2}
\end{minipage}

\begin{minipage}[t]{0.49\textwidth}
\includegraphics[width=\textwidth]{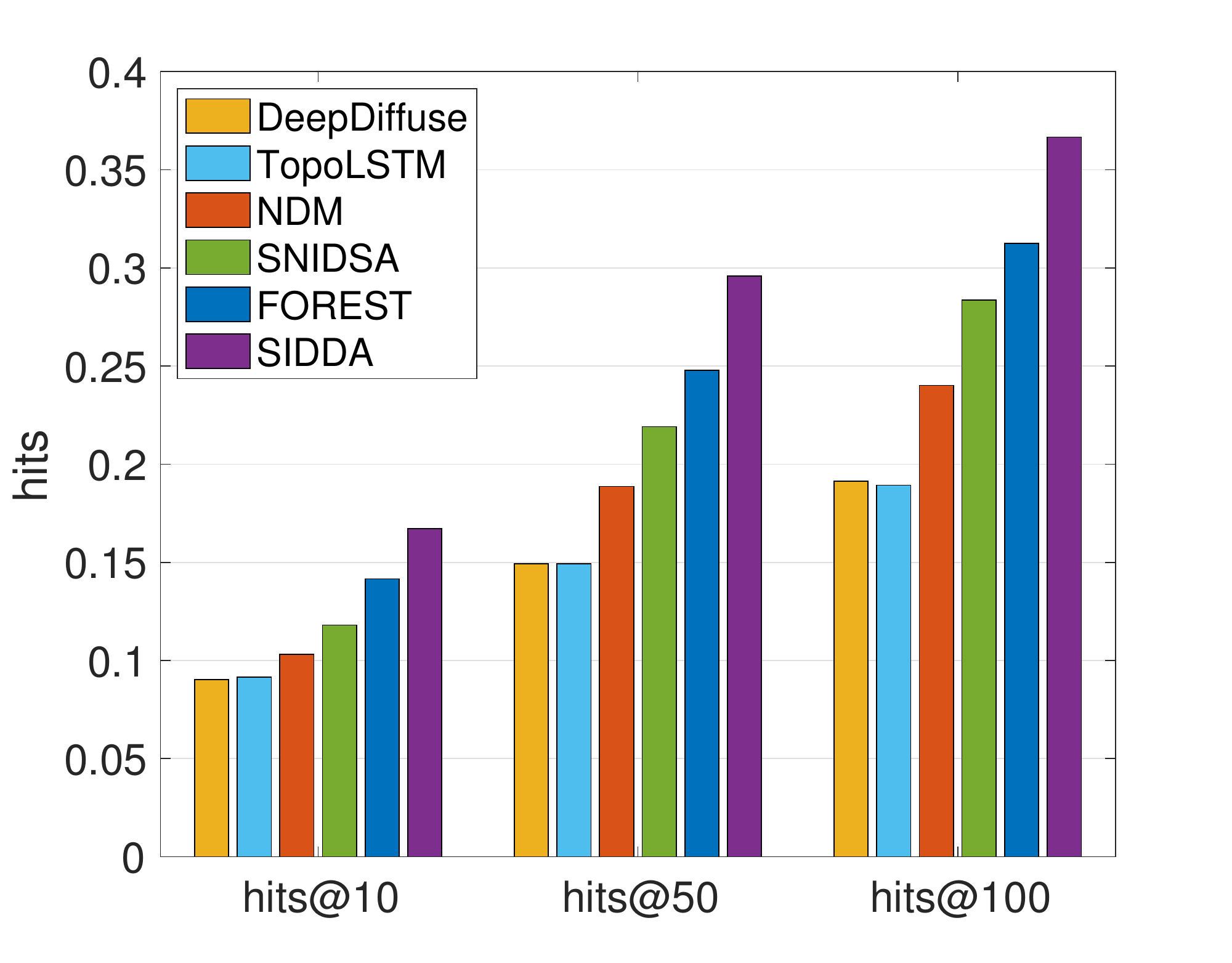}
\vspace*{-12mm}
\caption{$hits@N$ of Douban}
\label{overall3}
\end{minipage}
\begin{minipage}[t]{0.49\textwidth}
\includegraphics[width=\textwidth]{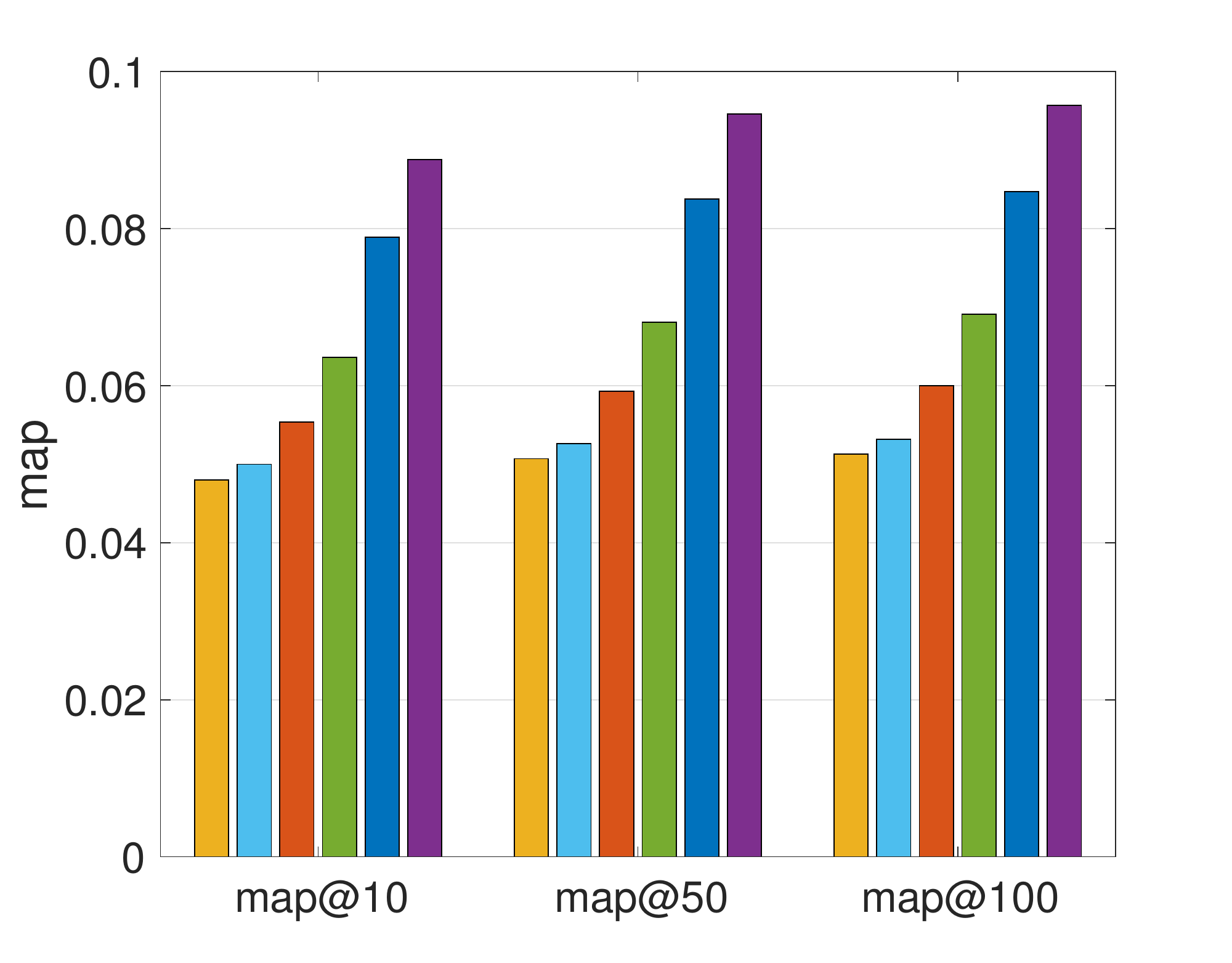}
\vspace*{-12mm}
\caption{$map@N$ of Douban}
\label{overall4}
\end{minipage}

\begin{minipage}[t]{0.49\textwidth}
\includegraphics[width=\textwidth]{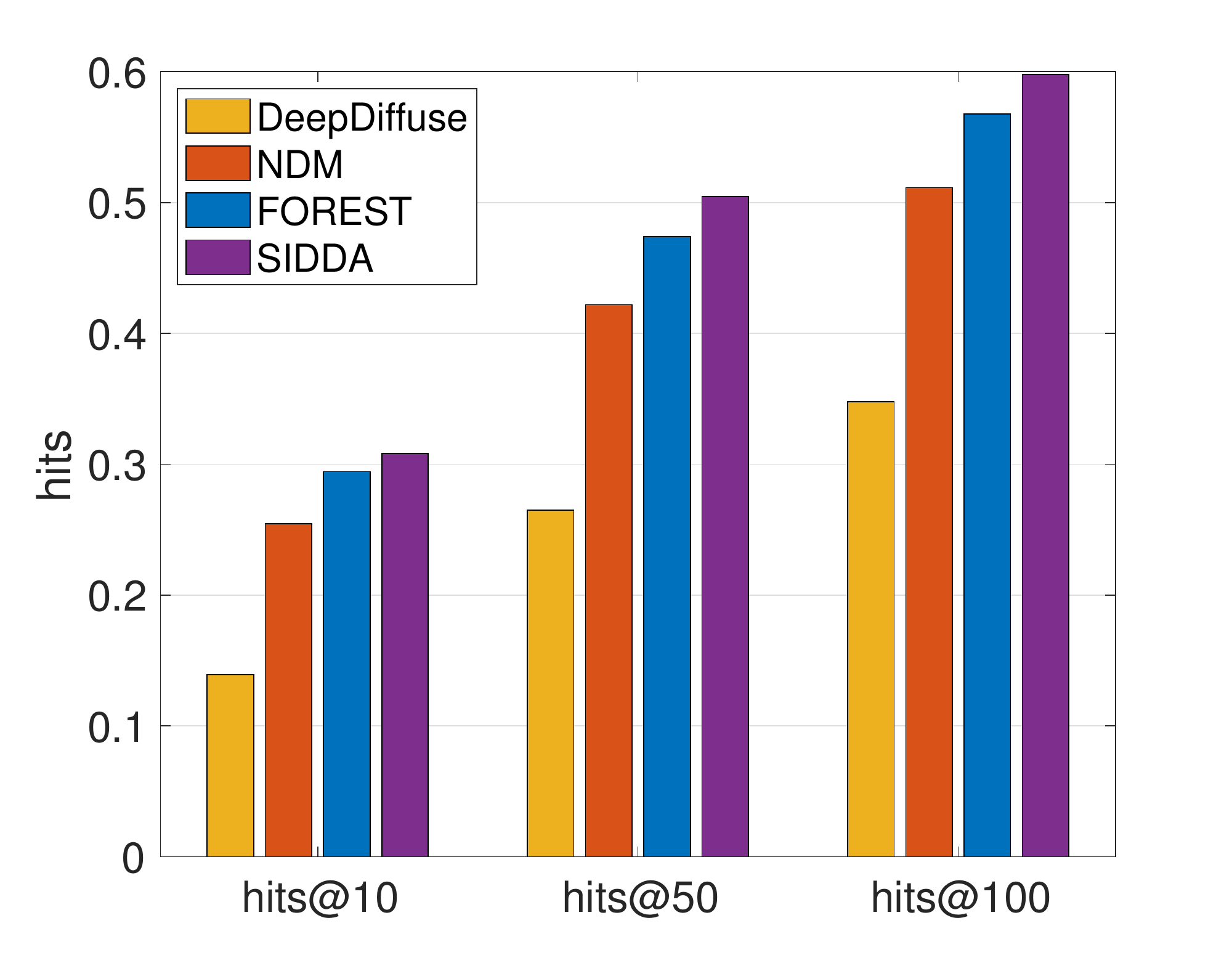}
\vspace*{-12mm}
\caption{$hits@N$ of Memetracker}
\label{overall5}
\end{minipage}
\begin{minipage}[t]{0.49\textwidth}
\includegraphics[width=\textwidth]{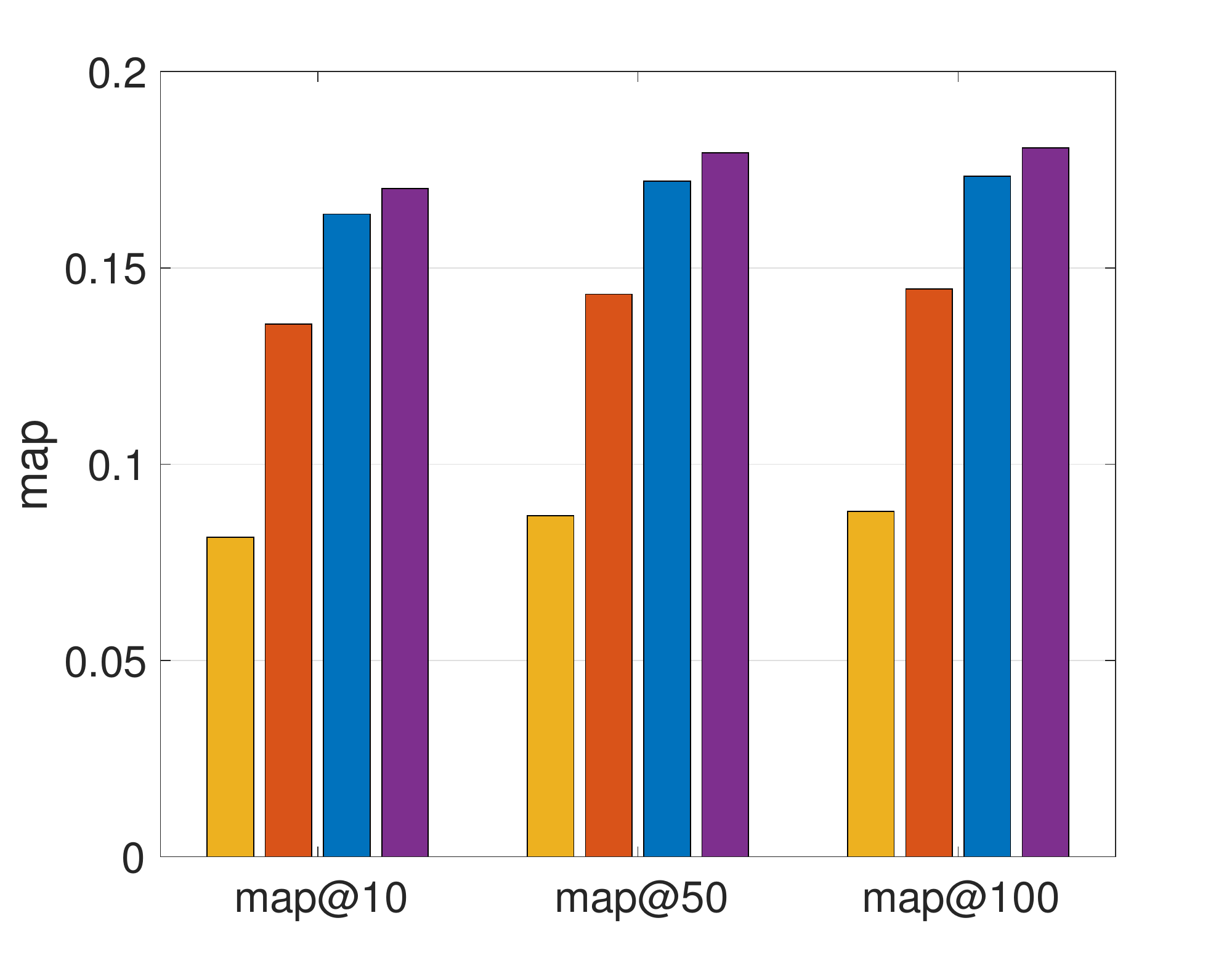}
\vspace*{-12mm}
\caption{$map@N$ of Memetracker}
\label{overall6}
\end{minipage}

\caption{The overall results, scores are the higher the better.}
\label{overall_result}
\end{figure}
\subsection{Experimental Results}
We design the comparative experiments, ablation experiments, and analyze the parameter sensitivity in this section.
\subsubsection{Overall Evaluation}
Fig. \ref{overall_result} shows the performance of different methods on three datasets. We select the results when $K=4$ for our method. For the analysis of parameters, please refer to Section \ref{Parameteranalysissec}. We find that:

SIDDA outperforms all the state-of-the-art baseline methods consistently and significantly on $hits@N$ and $map@N$. The results show that our method can predict the next activated user successfully.

\subsubsection{Ablation Study}
\label{albationstudysec}

\label{Parameteranalysissec}
\begin{figure}[htbp]
\centering
\includegraphics[width=\textwidth]{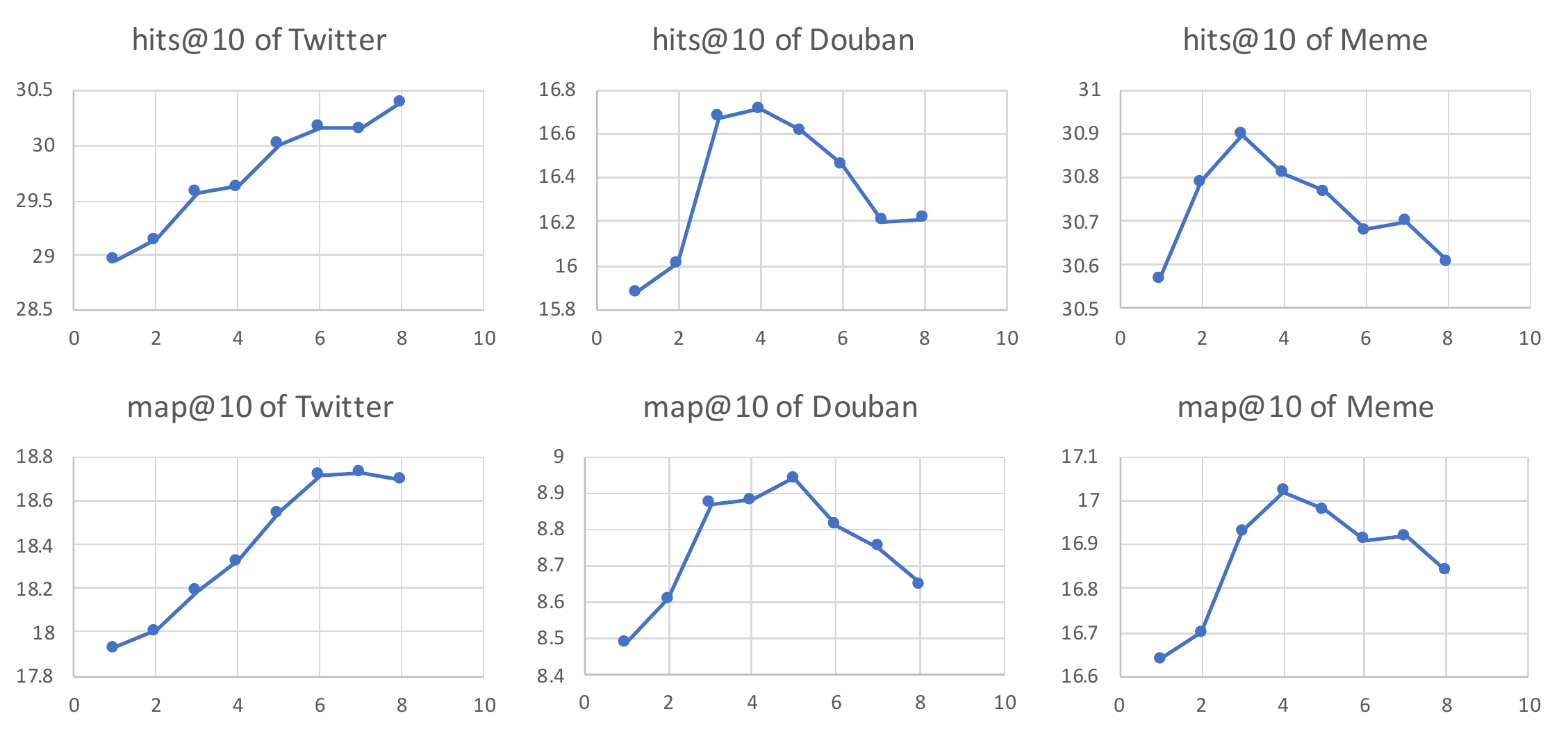}
\caption{Performance with different numbers of disentangled factors $K$.}
\label{figdifferentk}
\end{figure}

\paragraph{\textbf{The effect of the number of disentangled factors.}} We study how the number of disentangled factors $K$ can affect the performance of our method. As shown in Fig.\ref{figdifferentk}, our method has the worst performance when $K=1$. Because when $K=1$, our model is a GRU model with an attention mechanism, the latent information is not disentangled thoroughly. The performance on Douban and Memetracker dataset increases with $K$ getting larger, and start to decrease when $K>4$. While performance on the Twitter dataset is much better when $K>4$, showing an increasing trend. This is because Twitter has more diverse topics and may have more factors to disentangle. 

\begin{figure}[htbp]
\centering
\includegraphics[width=\textwidth]{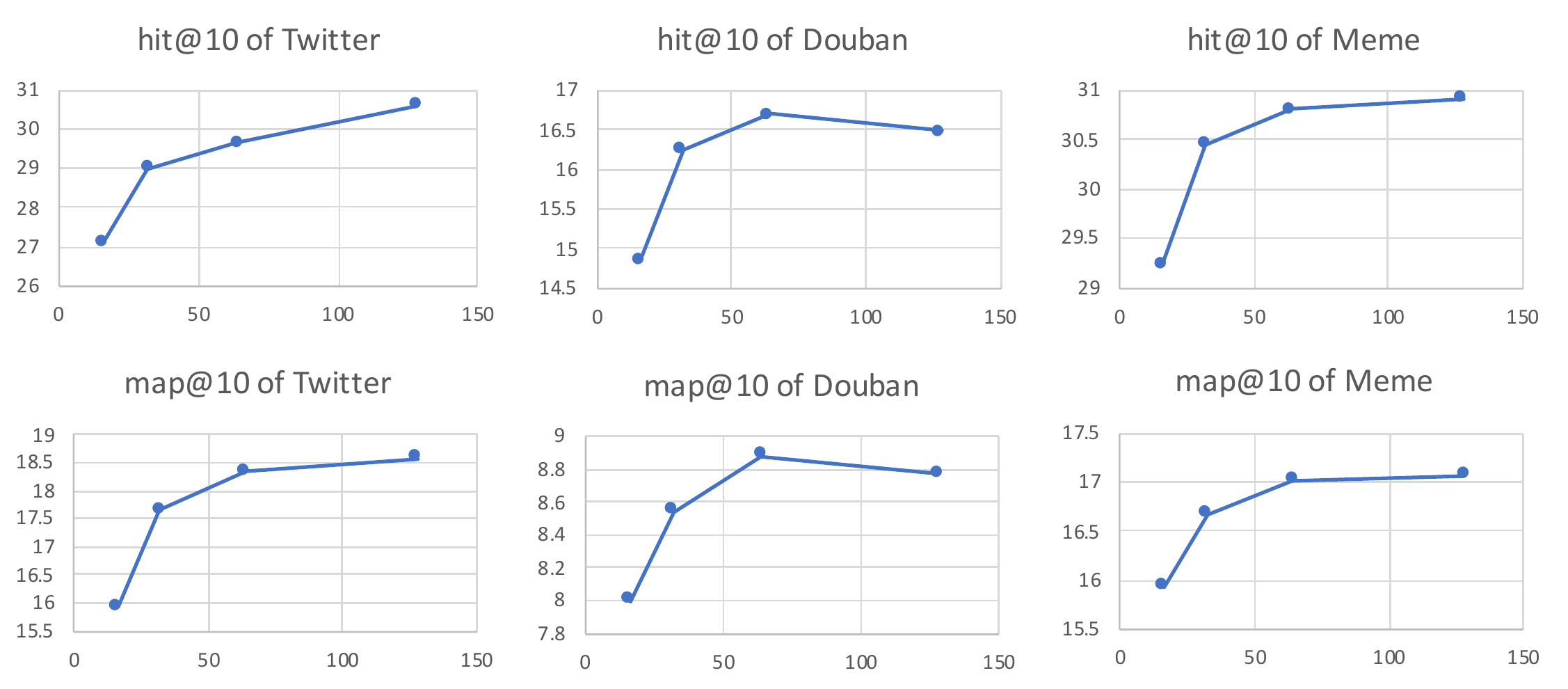}
\caption{Performance with different dimensions of node representations $D$.}
\label{figdifferentd}
\end{figure}

\paragraph{\textbf{The effect of the number of model dimension.}} We also study how the dimension of node representations $D$ can affect the performance. We evaluate our method when $K=4$ and $D\in\{16,32,64,128\}$. As illustrated in Fig. \ref{figdifferentd}, on Twitter and Memetracker the performance doesn't converge until $D=128$, possibly because they have larger training sets. Douban dataset converges when $D=64$ and shows a decreasing trend when $D$ gets larger.

\section{Conclusion}
In this paper, we propose a novel diffusion prediction model, SIDDA, which can extract latent factors by learning multiple hidden states at each timestamp. Specifically, we propose sequential attention to capture sequential relations of information cascades. Then we adopt a disentangled attention to learn multiple hidden states representing different latent factors. Additionally, we employ Gumbel-Softmax to further enhance the disentangled effect. The experiments on three real-world datasets show the superiority of our method when compared with state-of-the-art diffusion prediction models.

\bibliography{mybibfile}

\end{document}